\documentclass[twocolumn,amsmath,amssymb,superscriptaddress,aps,prc]{revtex4-1}

\usepackage{graphicx}
\usepackage{epsf}
\usepackage{amsmath,amssymb}
\usepackage{bm}
\usepackage{color}
\usepackage{ulem}

\begin{document}

\title{
Explicit inclusion of spin-orbit contribution in THSR wave function
}%

\author{N. Itagaki}

\affiliation{
Yukawa Institute for Theoretical Physics, Kyoto University,
Kitashirakawa Oiwake-Cho, Kyoto 606-8502, Japan
}

\author{H. Matsuno}

\affiliation{
Department of Physics, Kyoto University, Kitashirakawa Oiwake-Cho, Kyoto 606-8502, Japan
}

\author{A. Tohsaki}

\affiliation{
Research Center for Nuclear Physics,
10-1 Mihogaoka, Ibaraki, Osaka 067-0047,Japan
}

\date{\today}

\begin{abstract}
The Tohsaki-Horiuchi-Schuck-R\"{o}pke (THSR) wave function has been 
successfully used for the studies of gas-like nature of $\alpha$ clusters
in various nuclei including the so-called Hoyle state of $^{12}$C and four $\alpha$ states of $^{16}$O. 
In standard $\alpha$ cluster models, however,
each $\alpha$ cluster wave function has spin zero 
because of its spatial symmetry and antisymmetrization effect.
Thus the non-central interactions do not contribute, and
this situation is the same in the THSR wave function.
In this work, the spin-orbit contribution, which is found to be quite important at 
short $\alpha$-$\alpha$ distances, is taken into account 
in the THSR wave function by combing it with antisymmetrized quasi cluster model (AQCM).
The application to $^{12}$C is presented. 
The multi-integration in the original THSR wave function is carried out
by using Monte Carlo technique, which is called Monte Carlo THSR wave function.
For the nucleon-nucleon interaction, the Tohsaki interaction, which contains
finite-range three-body terms and simultaneously reproduces 
the saturation properties of nuclear systems, the $\alpha$-$\alpha$ scattering 
phase shift, and the size and binding energy of $^4$He, is adopted.
\end{abstract}

\pacs{21.30.Fe, 21.60.Cs, 21.60.Gx, 27.20.+n}
\maketitle

\section{Introduction}
The Tohsaki-Horiuchi-Schuck-R\"{o}pke (THSR) wave function has been 
widely used for the studies of gas-like states of $\alpha$ clusters,
including the so-called Hoyle state of $^{12}$C and four $\alpha$ states 
of $^{16}$O~\cite{PhysRevLett.87.192501,FUNAKI201578}.  
In normal $\alpha$ cluster models, each $\alpha$ cluster is described by
Gaussian-type wave function, and
the positions of the $\alpha$ clusters are specified by
Gaussian center parameters. 
In the THSR framework, all of these Gaussian center parameters
are integrated out with a weight function,
which enables us to describe well extended cluster states.
Also, by choosing a small range for the weight function,
the lowest configuration of the harmonic oscillator shell model
is also described.
Recently, ``container picture" has been proposed for 
the description of nonlocalized clusters~\cite{PhysRevLett.110.262501},
and also, the framework is applied even to the nuclei such as $^9$Be $etc.$, which do not
belong to
$4N$ nuclei~\cite{PhysRevC.91.014313}.

One of the problems of the traditional cluster models is that
the non-central interactions, especially the spin-orbit  interaction,
which is quite important in explaining the observed magic numbers,
do not contribute; they work neither inside $\alpha$ clusters nor between $\alpha$ clusters.
In cluster models, each $\alpha$ cluster is often defined as a simple $(0s)^4$ configuration
at some spatially localized point. In this case the antisymmetrization effect 
automatically makes the $\alpha$ cluster a spin singlet system free from
the non-central interactions.
To include the spin-orbit contribution
starting with the cluster model,
we proposed the antisymmetrized quasi cluster model
(AQCM)~\cite{PhysRevC.94.064324,PhysRevC.73.034310,PhysRevC.75.054309,PhysRevC.79.034308,PhysRevC.83.014302,PhysRevC.87.054334,ptep093D01,ptep063D01,ptepptx161,PhysRevC.97.014307}.
This method allows us to smoothly transform $\alpha$ cluster model wave functions to
$jj$-coupling shell model ones, and
we call the clusters that feel the spin-orbit effect after this treatment quasi clusters.
In AQCM,
we have only two parameters: $R$ (fm) representing the distance between $\alpha$ clusters
and $\Lambda$ $(-)$ characterizing the transition of $\alpha$ cluster(s) to quasi cluster(s).
It has been known that the conventional $\alpha$ cluster models cover the model space of closure of major shells
($N=2$, $N=8$, $N=20$, {\it etc.}) of the $jj$-coupling shell model.
In addition, we have shown that
the subclosure configurations of the $jj$-coupling shell model,
$p_{3/2}$ ($N=6$), $d_{5/2}$ ($N=14$), $f_{7/2}$ ($N=28$), and $g_{9/2}$ ($N=50$),
where spin-orbit effect is quite important,
can be reasonably described by our AQCM~\cite{ptep093D01}.

For such calculations, which include both cluster and shell features,
we inevitably  need a reliable nucleon-nucleon interaction, not cluster-cluster interaction.
It is quite well known that
the central part of the interaction 
should have proper density dependence in order
to satisfy the saturation property of nuclear systems.
If we just introduce simple
two-body interaction, for instance Volkov interaction~\cite{VOLKOV196533}, which has been
widely used in the cluster studies, we have to
properly choose Majorana exchange parameter for each nucleus, and consistent description of
two different nuclei with the same Hamiltonian becomes almost impossible.
Adding zero-range three-body interaction term helps better agreements with
experiments; however
the radius and binding energy of free $^4$He ($\alpha$ cluster) are not well 
reproduced~\cite{PTP.64.1608}.
The Tohsaki interaction, which has finite range three-body terms, 
has much advantages~\cite{PhysRevC.49.1814,PhysRevC.94.064324,PTP.94.1019}.
Although this is a phenomenological interaction,
it gives reasonable size and binding energy of the $\alpha$ cluster,
and $\alpha$-$\alpha$ scattering phase shift is reproduced,
while
the saturation properties of nuclear matter is also reproduced rather sufficiently.

In this paper, we combine the THSR wave function and the idea of AQCM.
It is worthwhile to show the applicability of the combined method 
by numerical calculations. The THSR wave function contains the multi-integration 
over the Gaussian center parameters of $\alpha$ clusters, and this procedure 
can be numerically performed using Monte Carlo technique
called Monte Carlo THSR wave function~\cite{PhysRevC.75.037303,PhysRevC.77.037301,PhysRevC.78.017306,PhysRevC.80.021304,PhysRevC.82.014312}.
In the present study, we 
calculate $^{12}$C (three $\alpha$) as the first step.

\section{Framework}

In this section, we summarize the essence of the THSR wave function and AQCM,
and combination of these two is newly introduced.

\subsection{Brink model}

The THSR wave function is based on the Brink model~\cite{Brink}. 
Each single particle wave function
of the Brink model is described by a Gaussian,
\begin{equation}	
	\phi_{ij} = \left(  \frac{2\nu}{ \pi } \right)^{\frac{3}{4}} 
		\exp \left[-  \nu \left(\bm{r}_{i} - \bm{R}_{i} \right)^{2} \right] \chi_{j}, 
\label{spwf} 
\end{equation}
where
the Gaussian center parameter $\bm{R}_{i}$ shows the expectation 
value of the position of the $i$-th $\alpha$ cluster.
The index $j$ specifies four nucleons in this $i$-th $\alpha$ cluster,
and $\chi_{j}$ represents the spin isospin part
of the single particle wave function.
The size parameter $\nu$ is chosen to be 0.25~fm$^{-2}$,
which reproduces the observed radius of $^4$He.

The Slater determinant of the Brink model is constructed from 
these single particle wave functions by antisymmetrizing them. 
Here, four single particle 
wave functions with different spin and isospin
sharing a common Gaussian center parameter
correspond to an $\alpha$ cluster.

\begin{eqnarray}
\Phi_{SD}(\bm{R}_1, \bm{R}_2, \ldots, \bm{R}_N)
= &&  
{\cal A} \{ (\phi_{11}\phi_{12}\phi_{13}\phi_{14})  (\phi_{21}\phi_{22}\phi_{23}\phi_{24})   \nonumber \\
&& \ldots (\phi_{N1}\phi_{N2}\phi_{N3}\phi_{N4}) \}.
\label{SD} 
\end{eqnarray}
This is the case that we have $N$ $\alpha$ clusters and the mass number $A$ is equal to
$A = 4N$.

\subsection{THSR wave function}
The idea of the THSR wave function~\cite{PhysRevLett.87.192501} is that Gaussian center parameters
$\{ \bm{R}_{i}  \}$ 
are integrated over infinite space with the weight functions
$\{ \exp[-\bm{R}_{i}^2 / \sigma^2] \}$.
Thus the THSR wave function $\Phi_{THSR}$ 
is expressed using 
$\Phi_{SD}$ in Eq.~\eqref{SD} as
\begin{eqnarray}
\Phi_{THSR} = && \int d\bm{R}_1 d\bm{R}_2 \cdots d\bm{R}_N
\ \Phi_{SD}(\bm{R}_1, \bm{R}_2, \ldots, \bm{R}_N)
\nonumber \\ 
&& 
\times \exp[ -(\bm{R}_1^2+\bm{R}_2^2 \cdots +\bm{R}_N^2)  / \sigma^2 ].
\label{THSR}
\end{eqnarray}
Here $\sigma$
is a control parameter, which governs the
spatial extension of the system.
When $\sigma$ is large, the wave function describes
gas-like states of $\alpha$ clusters, and
the lowest configuration of the harmonic oscillator shell model
can be realized at the limit of $\sigma \to 0$.

\subsection{Monte Carlo THSR}
In some cases of the original THSR wave function,
the analytic formula for the matrix elements 
for the Hamiltonian is already obtained.
However for heavier nuclei, it is useful to introduce Monte Carlo technique for 
the multi-integration in the original THSR wave 
function~\cite{PhysRevC.75.037303,PhysRevC.77.037301,PhysRevC.78.017306,PhysRevC.80.021304,PhysRevC.82.014312}.
We call this wave function ($\Phi_{M-THSR}$) Monte Carlo THSR,
\begin{equation}
\Phi_{M-THSR} = \sum_k^{N_{max}} P^{J^\pi} \Phi_{SD}^k(\bm{R}_1, \bm{R}_2, \ldots, \bm{R}_N).
\label{vTHSR}
\end{equation}
Here the multi-integration over the Gaussian center parameters 
in Eq.~\eqref{THSR}
is replaced with a summation of different Slater determinants.
The Slater determinants superposed have different values 
of Gaussian center parameters 
$\{\bm{R}_i\}^k$ for $N$ $\alpha$ clusters, where $k$ is a number 
to specify the set of the Gaussian center parameters
for the $k$-th Slater determinant. 
The value of the Gaussian center parameters are randomly generated, but the distribution of 
the absolute value $|\bm{R}_i|$
for the $i$-th $\alpha$ cluster is introduced to be proportional to 
$\exp[ -\bm{R}_i^2/\sigma^2]$, and
its angular part is isotropically generated.
Thus the information of the weight function 
in the original THSR wave function
is absorbed in the distribution of randomly generated $\{\bm{R}_i\}^k$ values.
The value of $N_{max}$ shows the number of Slater determinants, which are superposed.
The limit of $N_{max} \to \infty$ coincides with the original THSR wave function;
however we approximate it with a finite number.
In Eq.~\eqref{vTHSR}, $P^{J^\pi}$ shows the projection onto the eigen states
of parity and angular momentum, and this is numerically performed.

\subsection{AQCM}

In the conventional cluster models,
there is no spin-orbit effect for the $\alpha$ clusters.
Thus
they are changed into quasi clusters based on 
AQCM~\cite{PhysRevC.94.064324,PhysRevC.73.034310,PhysRevC.75.054309,
PhysRevC.79.034308,PhysRevC.83.014302,PhysRevC.87.054334,
ptep093D01,ptep063D01,ptepptx161,PhysRevC.97.014307}.
According to AQCM,
when the original position of one of the $\alpha$ clusters 
(the value of Gaussian center parameter) is $\bm{R}$,
the Gaussian center parameter of the $l$-th nucleon in this cluster is transformed 
by adding the imaginary part as
\begin{equation}
\bm{\zeta}_l = \bm{R} + i \Lambda \bm{e}^{\text{spin}}_l \times \bm{R}, 
\label{AQCM}
\end{equation}
where $\bm{e}^{\text{spin}}_l$ is a unit vector for the intrinsic-spin orientation of the 
$l$-th nucleon in this $\alpha$ cluster.
For the imaginary part,
here we introduce $\Lambda$, which is a real dimensionless parameter.
After this transformation, this $\alpha$ cluster is called quasi cluster.
The imaginary part, which is added, depends on the spin direction of each nucleon,
thus the $\alpha$ cluster is no longer a spin singlet system.
The spin-orbit contribution can be taken into account by this transformation,
and the contribution is attractive when $\Lambda$ is positive.
We have previously shown that
the lowest configurations of the $jj$-coupling shell model
can be achieved by $\Lambda = 1$ and $\bm{R} \to 0$.

\subsection{Monte Carlo THSR+AQCM}

We propose a new framework by combining Monte Carlo THSR and AQCM,
which is applied to $^{12}$C.
For $^{12}$C, we introduce AQCM for all the three $\alpha$ clusters.
Here the intrinsic-spins of the nucleons  in the three $\alpha$ clusters 
are introduced to have threefold symmetry.
This is needed to include the lowest configuration of the $jj$-coupling 
shell model (subclosure configuration of $p_{3/2}$) 
within a single Slater determinant.
In the first $\alpha$ cluster, each intrinsic-spin of the four nucleons
is spin-up ($z$ direction) for a proton and a neutron
and spin-down ($-z$ direction) for a proton and a neutron.
The intrinsic-spin orientations of the four nucleons in the second 
and third $\alpha$ clusters are introduced by rotating the ones
of the first $\alpha$ cluster by $2\pi/3$ and $4\pi/3$ radians, respectively.
These spin orientations of the twelve nucleons are fixed 
in all the Slater determinant before the angular momentum projection.
While fixing the intrinsic-spin orientations,
at first we randomly generate the Gaussian center parameters 
of three $\alpha$ clusters based on Monte Carlo THSR,
and next, the center of mass of the total is shifted to the origin.
Finally, 
imaginary parts of the Gaussian center parameters are introduced based on AQCM
as in Eq.~\eqref{AQCM},
and angular momentum projection and superposition of different 
Slater determinants follow.
Here the second step of shifting the center of mass to the origin is quite important;
the purpose of AQCM treatment is to describe the single particle motion of each nucleon
around the origin and take into account the spin-orbit contribution, 
thus the center of the nucleus has to coincide 
with the origin of the coordinate system before giving the imaginary part.

\subsection{Hamiltonian}

The Hamiltonian ($\hat{H}$) consists of kinetic energy ($\hat{T}$) and 
potential energy ($\hat{V}$) terms,
and the kinetic energy term is described as one-body operator,
\begin{equation}
\hat{T} = \sum_i \hat{t_i} - T_{cm},
\end{equation}
and the center of mass kinetic energy ($T_{cm}$),
which is constant,
is subtracted.
The potential energy has
central, spin-orbit, 
and the Coulomb parts.
For the central part, 
we introduce the Tohsaki interaction~\cite{PhysRevC.49.1814}, which has
finite range three-body terms in addition to the two-body nucleon-nucleon interaction terms.
This interaction is designed to reproduce both saturation property and
the scattering phase shift of two $\alpha$ clusters.
We use the F1 parameter set~\cite{PhysRevC.49.1814}
in the present analysis,
which was used in the original THSR work for $^{12}$C.

For the spin-orbit part ($\hat{V}_{so}$),
the spin-orbit term of G3RS \cite{PTP.39.91}, which is a realistic
interaction originally determined to reproduce the nucleon-nucleon scattering phase shift, 
is adopted;
\begin{equation}
\hat{V}_{so}= \sum_{i < j} 
V_{ls}( e^{-d_{1} (\vec r_i - \vec r_j)^{2}}
                    -e^{-d_{2} (\vec r_i - \vec r_j)^{2}}) 
                     P(^{3}O){\vec{L}}\cdot{\vec{S}},
\label{Vls}
\end{equation}
 where $d_{1}= 5. 0$~fm$^{-2},\ d_{2}= 2. 778$~fm$^{-2}$,
and $P(^{3}O)$ is a projection operator onto a triplet odd state.
The operator $\vec{L}$ stands for the relative angular momentum
 and $\vec{S}$ is the total spin, ($\vec{S} = \vec{S_{1}}+\vec{S_{2}}$).
The strength,
$V_{ls}$, has been determined to reproduce the $^4$He+$n$
scattering phase shift~ \cite{PTP.57.866}, and
$V_{ls} = 1600-2000$~MeV has been suggested.
Here we employ $V_{ls} = 1800$~MeV, which has been tested 
in our previous works for $^{12}$C~\cite{PhysRevC.94.064324},
although there the Majorana parameter for the three-body (central) interaction
is slightly modified to reproduce the binding energy of $^{16}$O.

\section{Results and discussions}

\begin{figure}[htbp]
	\centering
	\includegraphics[width=6.5cm]{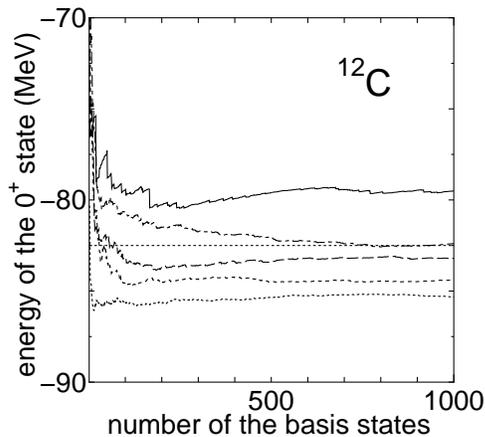} 
	\caption{
The $0^+$ energy convergence of $^{12}$C (three $\alpha$'s)
as function of the number of Slater determinants superposed
 ($k$ in Eq.~\eqref{vTHSR})
calculated with the Monte Carlo THSR framework.
The $\alpha$ clusters are not broken ($\Lambda = 0$ in Eq.~\eqref{AQCM}), and
the solid, dotted, dashed, short-dashed, dashed, and dash-dotted lines correspond to 
$\sigma = 1$, 2, 3, 4, and 5~fm in Eq.~\eqref{THSR}, respectively.
The dashed line at $-82.50$~MeV shows the three $\alpha$ threshold energy.
     }
\label{toh-brink}
\end{figure}

We start Monte Carlo THSR calculation with the $\alpha$ non-breaking case,
which is nothing but the superposition of Brink-type $\alpha$ cluster model
wave functions.
In the AQCM framework, this situation corresponds to 
setting $\Lambda$ in Eq.~\eqref{AQCM} to zero.
In Fig.~\ref{toh-brink},
the $0^+$ energy convergence of $^{12}$C (three $\alpha$ clusters) is shown
as a number of Slater determinants superposed
based on the Monte Carlo THSR framework ($k$ in Eq.~\eqref{vTHSR}).
The solid, dotted, short-dashed, dashed, and dash-dotted lines correspond to 
$\sigma = 1$, 2, 3, 4, and 5~fm in Eq.~\eqref{THSR}, respectively.
The dashed line at $-82.50$~MeV shows the three $\alpha$ threshold energy.
This is Monte Carlo calculation and not variational one,
thus the energy is not always going down; the energy sometimes goes up
with increasing number of the basis states.
Nevertheless,
the energy converges at the limit of $N_{mas} \to \infty$,
but here we can confirm that it is well converged with 1000 basis states
($N_{max}=1000$ in Eq.~\eqref{vTHSR}).
In this calculation, the radial part of 
the Gaussian center parameters 
$\{\bm{R}_i\}^k$
are generated by the random numbers
${\{ r_i \}}$,
whose distribution is proportional to $\exp[-r_i^2/\sigma^2]$, and
the angular part of each $\bm{R}_i$ is isotropically generated using random numbers.
It is found that the converged energy 
of small $\sigma$ case, $\sigma=1$~fm (solid line), is above the three $\alpha$ threshold.
This is because the contribution of the spin-orbit interaction, which is important in inner regions,
is missing within the $\Lambda = 0$ wave functions. 
Other $\sigma$ values give the energies bellow the threshold,
and the dotted line ($\sigma = 2$~fm) gives the lowest energy. 
With increasing the $\sigma$ value, the energy again goes up,
and the energy of $\sigma = 5$~fm (dash-dotted line) is close to the threshold.

\begin{figure}[htbp]
	\centering
	\includegraphics[width=6.5cm]{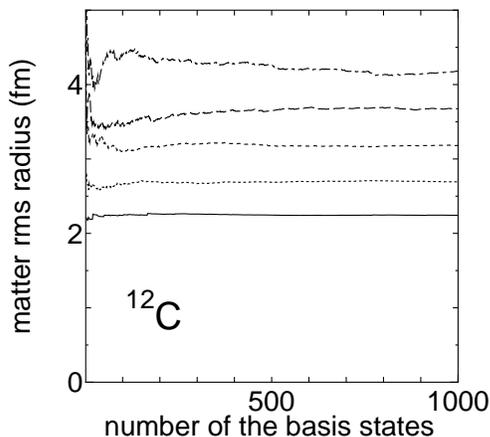} 
	\caption{
The convergence of the root mean square (rms) matter radius for the $0^+$ state of $^{12}$C
as a function of the number of Slater determinants superposed
 ($k$ in Eq.~\eqref{vTHSR})
calculated with the Monte Carlo THSR framework.
The $\alpha$ clusters are not broken ($\Lambda$ in Eq.~\eqref{AQCM} is zero), and
the solid, dotted, dashed, short-dashed, dashed, and dash-dotted lines correspond to 
$\sigma = 1$, 2, 3, 4, and 5~fm in Eq.~\eqref{THSR}, respectively.
}
\label{mrms}
\end{figure}

\begin{figure}[htbp]
	\centering
	\includegraphics[width=6.5cm]{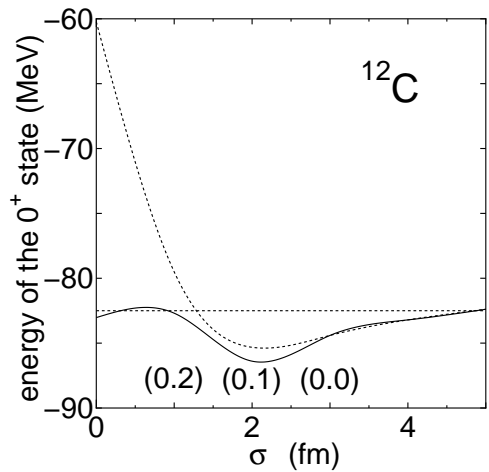} 
	\caption{
The $0^+$ energy curve of $^{12}$C as a function of
$\sigma$ (fm) defined in Eq.~\eqref{THSR}.
The $\Lambda$ values
defined in Eq.~\eqref{AQCM} is a variational parameter
in the solid line, whereas $\Lambda$ is fixed to zero in the dotted line.
For the solid line,
the values in the parentheses show the optimal $\Lambda$ values
for the cases of $\sigma =$ 1.0, 2.0, and 3.0~fm.
}
\label{toh-ec}
\end{figure}

\begin{figure}[htbp]
	\centering
	\includegraphics[width=6.5cm]{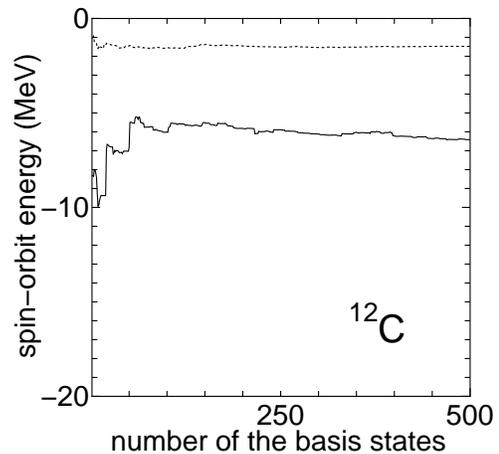} 
	\caption{
The convergence of the spin-orbit energy  for the $0^+$ state of $^{12}$C
as a function of 
number of Slater determinants superposed
 ($k$ in Eq.~\eqref{vTHSR}).
The solid line is for $\sigma =$ 1.0~fm and $\Lambda = 0.2$,
and the dotted line is for 
$\sigma =$ 2.0~fm and $\Lambda = 0.1$.
}
\label{toh-ls-c}
\end{figure}

Using these wave functions,
the convergence of the root mean square (rms) matter radius 
for the $0^+$ states of $^{12}$C
is shown in Fig.~\ref{mrms}.
The basis states are the same as those in Fig.~\ref{toh-brink},
and wave functions are $\Lambda = 0$ (Brink $\alpha$ cluster model), 
and the types of the lines are also the same;
the solid, dotted, short-dashed, dashed, and dash-dotted lines correspond to 
$\sigma = 1$, 2, 3, 4 and 5~fm in Eq.~\eqref{THSR}, respectively.
Experimentally the rms matter radius of $^{12}$C is obtained as 2.35(2)~fm in Ref.~\cite{PhysRevLett.117.102501},
consistent to the value deduced from the electron scattering,
and the rms radius of $\sigma = 1$~fm (solid line) is close to this value.
If we compare the energy in  Fig.~\ref{toh-brink} and the rms radius in Fig.~\ref{mrms},
we find that this interaction gives the optimal state with slightly larger rms radius
than experiments.
However the energy curve with respect to the $\sigma$ value drastically changes
if we incorporate the spin-orbit effect as we discuss shortly.
For the second $0^+$ state known as the Hoyle state,
the large rms radius of $\sim$4.0~fm has been extensively discussed,
although this state is a resonance state slightly above the threshold
and the experimental determination is difficult.
In the present case, the $\sigma = 5$~fm result (dash-dotted line) gives the energy around
the threshold and the rms radius of 4.18~fm.

Then we take finite $\Lambda$ values, which allows us to take into account
the spin-orbit contribution.
The $0^+$ energy curve for $^{12}$C is shown in Fig.~\ref{toh-ec} as a function of
$\sigma$ (fm) defined in Eq.~\eqref{THSR}.
The $\Lambda$ values
defined in Eq.~\eqref{AQCM} is a variational parameter
in the solid line, whereas $\Lambda$ is fixed to zero in the dotted line.
For the solid line,
the values in the parentheses show the optimal
$\Lambda$ values for the cases of $\sigma =$ 1.0, 2.0, and 3.0~fm.
We can see large decrease of the energy by more than 20~MeV at the limit of $\sigma = 0$~fm
after optimizing the $\Lambda$ value.
Then in the case of the solid line, a local minimum point appears
at the limit of $\sigma = 0$~fm, where the contribution of the spin-orbit interaction is very large,
more than $-30$~MeV.
The optimal $\Lambda$ value of 0.3 at the limit of $\sigma = 0$~fm shows that $\alpha$ clusters
are broken to some extent and the wave function approaches to the $jj$-coupling shell model one.
In general,
the contribution of the spin-orbit interaction increases and the energy decreases 
with increasing $\Lambda$ (for each fixed $\sigma$), but it saturates at some point.
On the other hand, the kinetic energy quadratically increases with increasing $\Lambda$,
and the energy minimum state appears owing to the compensation of these two factors.

As a function of $\sigma$, in Fig.~\ref{toh-ec},
the energy minimum point of the solid line appears around $\sigma = 2.0$~fm,
where the optimal $\Lambda$ value is 0.1.
Owing to the additional attraction of the spin-orbit interaction,
here the solid line is lower than the dotted line by about 1~MeV.
If we mix two minimum points, the true minimum point of ($\sigma$~fm,~$\Lambda$)~=~(2.0~fm,~0.1) and 
the local minimum point of ($\sigma$~fm,~$\Lambda$)~=~(0.0~fm,~0.3),
with the amplitude ratio of 2:1 after normalizing each of these two,
the rms matter radius becomes 2.37~fm, which reproduces the experimental value.

The convergence of the spin-orbit energy  for the $0^+$ state of $^{12}$C
is shown in Fig.~\ref{toh-ls-c}
as a function of 
number of Slater determinants superposed
 ($k$ in Eq.~\eqref{vTHSR}).
This is a demonstration that the spin-orbit effect can be successfully taken into account
with the procedure proposed here.
The solid line is for $\sigma =$ 1.0~fm and $\Lambda = 0.2$,
and the dotted line is for 
$\sigma =$ 2.0~fm and $\Lambda = 0.1$.
The $\Lambda$ values are optimal ones in each $\sigma$ case,
and the latter gives the lowest energy point in Fig.~\ref{toh-ls-c}.
The solid line ($\sigma$~fm,~$\Lambda$)~=~(1.0~fm,~0.2) converges to
$\sim -6.4$~MeV, whereas
the dotted line ($\sigma$~fm,~$\Lambda$)~=~(2.0~fm,~0.1) converges to
$\sim -1.5$~MeV.

\begin{figure}[htbp]
	\centering
	\includegraphics[width=6.5cm]{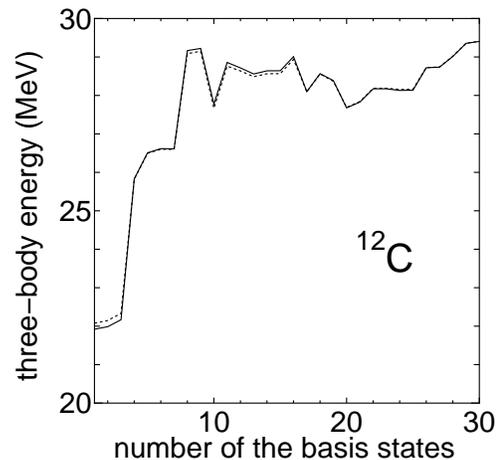} 
	\caption{
The contribution of three-body interaction terms for the $0^+$ state of $^{12}$C
as a function of the number of Slater determinants superposed
 ($k$ in Eq.~\eqref{vTHSR})
calculated with the Monte Carlo THSR framework.
The $\sigma$ (fm) value defined in Eq.~\eqref{THSR} is 2.0~fm, and
$\Lambda$ in Eq.~\eqref{AQCM} is 0.1 (solid line) and 0 (dotted line).
}
\label{three-body}
\end{figure}

\begin{figure}[htbp]
	\centering
	\includegraphics[width=6.5cm]{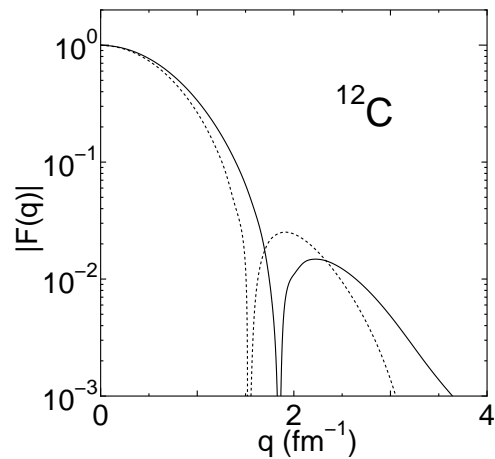} 
	\caption{
The absolute value of the elastic form factor ($|F_(q)|$) for the $0^+$ state of $^{12}$C
as a function of the momentum transfer $q$~(fm$^{-1}$).
The dotted line is for $(\sigma$~fm,~$\Lambda)$~=~(2.0~fm,~0.1), which gives the optimal
energy as shown in Fig.~\ref{toh-ec}, and the solid line is the result after mixing
the local minimum point (at the limit of $\sigma=0$~fm limit,~$\Lambda=0.3$) with the amplitude ratio of 2:1.
The experimental values and other theoretical results are compared 
in Refs.~\cite{PTPS.68.29,PhysRevLett.111.092501}.
}
\label{c12ff}
\end{figure}

After introducing the finite value of the $\Lambda$ values,
each nucleon is more independently treated and calculation costs
significantly increases compared with the Brink model ($\Lambda = 0$) calculation.
Therefore, in Fig.~\ref{toh-ec}, although the contribution of
the kinetic energy is calculated with superposing 1000 Slater determinants
($N_{max}=1000$ in Eq.~\eqref{vTHSR}),
the contribution of 
the two-body interactions is estimated with 500 Slater determinants
($N_{max} = 500$).
The most time consuming part is the finite-range three-body interaction part.
This three-body part is substituted with the values obtained with
the $\Lambda =0$ wave functions. The three-body interaction
terms do not strongly depend on the $\Lambda$ values, and we can approximate it
with the Brink model. The Brink model calculation is rather simple and we can superpose
1000 basis states for the estimation of the three-body terms.
This approximation can be justified in Fig.~\ref{three-body},
which shows the contribution of the three-body interaction.
The $\sigma$ value is 2.0~fm,  and the solid line is
for $\Lambda = 0.1$, which gives the optimal energy in Fig.~\ref{toh-ec}, and the  dotted line is for  $\Lambda = 0.0$.
The real parts of Gaussian center parameters for each Slater determinant 
are common for the $\Lambda = 0.0$ (dotted line) and $\Lambda = 0.1$ (solid line) cases,
and imaginary parts are just added to the real parts in the $\Lambda = 0.1$ case following Eq.~\eqref{AQCM}.
The energies are not converged yet within such a small number of the basis states,
but the values of these two lines for the three-body interaction terms  are very close
not to be distinguished.
Indeed, the difference is less than~100 keV.
Then we can estimate 
the contribution of three-body terms in the finite $\Lambda$ cases
by superposing Brink-type Slater determinants ($\Lambda = 0$), 
where number of the basis states ($N_{max}$) is not 30 as in this figure but 
increased to 1000.

The absolute value of the elastic form factor ($|F_(q)|$) for the $0^+$ state of $^{12}$C
is shown in Fig.~\ref{c12ff}
as a function of the momentum transfer $q$~(fm$^{-1}$).
The dotted line is for ($\sigma$~fm,~$\Lambda$)~=~(2.0~fm,~0.1), which gives the optimal
energy as shown in Fig.~\ref{toh-ec}. The solid line is the result after mixing
the local minimum point (at the limit of $\sigma=0$ fm, $\Lambda = 0.3$). The mixing ratio of 
($\sigma=2.0$~fm,~$\Lambda = 0.1$) : ($\sigma=0$~fm,~$\Lambda = 0.3$) is given as 2:1 after normalizing each configuration.
The experimental values and other theoretical results are compared 
in Refs.~\cite{PTPS.68.29,PhysRevLett.111.092501}.
The dotted line line shows the sign change around $q\sim 1.6$~fm$^{-1}$,
which is too small compared with the experiment, reflecting the fact that
the ($\sigma$~fm,~$\Lambda$)~=~(2.0~fm,~0.1)
configuration, which is energetically optimal one,
has too large spacial extension compared with the experiment
(experimentally this sign change occurs around $q \sim 1.8$~fm$^{-1}$).
We have previously mentioned that the mixing of two states, the energy optimal one and the local minimum point
with smaller radius,
enables us to reproduce the experimental rms radius,
and this mixing turns out to be also important in reproducing the form factor, which is the solid line.

\section{Summary}\label{summary}
The THSR wave function has been 
successfully used for the studies of gas-like nature of $\alpha$ clusters
of various nuclei.
In this work, we proposed a method to take into account
the spin-orbit contribution 
in THSR by combing it with AQCM.
In the standard $\alpha$ cluster models, 
each $\alpha$ cluster wave function has spin zero 
because of the spatial symmetry of the $\alpha$ clusters and antisymmetrization effect.
Thus the non-central interactions do not contribute, and
this situation is the same in the THSR wave function.
The application of a new framework to $^{12}$C was presented. 
The multi-integration in the original THSR wave function was carried out
by using Monte Carlo technique, 
which is called Monte Carlo THSR wave function.
 In $^{12}$C,
the contribution of the spin-orbit interaction 
was successfully taken into account.
Especially for the cases when
the spatial extension is small, the contribution is quite strong,
but it decreases
with increasing the spatial extension.
As a result, one local minimum at the limit of zero distance between $\alpha$ clusters
and the real minimum state with sizable $\alpha$-$\alpha$ distances appear.
If we mix these two configurations, we can reproduce the
observed matter rms radius.
This can be considered as the quantum mechanical mixing of different structures,
or more concretely, competition of shell and cluster structures.

As a future work, we further apply the combined framework of
Monte Carlo THSR and AQCM, which was proposed in the present study, to other cases. 
Also, we try to derive analytic formula for the matrix elements 
of the Hamiltonian for the combined framework.

\begin{acknowledgments}
Numerical calculation has been performed using the computer facility of 
Yukawa Institute for Theoretical Physics,
Kyoto University. This work was supported by JSPS KAKENHI Grant Number 17K05440.
\end{acknowledgments}

\bibliographystyle{apsrev4-1}
\bibliography{biblio_ni.bib}

\end{document}